\documentclass[iop]{emulateapj}

\usepackage{amsmath}
\newcommand{\angstrom}{\text{\normalfont\AA}}
\shorttitle{A Ground-based Optical Transmission Spectrum of WASP-6b}
\shortauthors{Jord\'an et al. 2013}

\usepackage{comment}

\slugcomment{Accepted for publication in the Astrophysical Journal}
\begin{document}

\title{A Ground-based Optical Transmission Spectrum of WASP-6\MakeLowercase{b}}

\author{Andr\'es Jord\'an\altaffilmark{1}, 
N\'estor Espinoza\altaffilmark{1}, 
Markus Rabus\altaffilmark{1}, 
Susana Eyheramendy\altaffilmark{2},
David K. Sing\altaffilmark{3}, 
Jean-Michel D\'esert\altaffilmark{4,5}, 
G\'asp\'ar  \'A. Bakos\altaffilmark{6,11,12}, 
Jonathan J. Fortney\altaffilmark{7}, 
Mercedes  L\'opez-Morales\altaffilmark{8}, 
Pierre F.L. Maxted\altaffilmark{9},
Amaury H.M.J. Triaud\altaffilmark{10,13}, 
Andrew Szentgyorgyi\altaffilmark{8}
}

\altaffiltext{1}{Instituto de Astrof\'isica, Facultad de F\'isica, Pontificia Universidad
  Cat\'olica de Chile, Av.\ Vicu\~na Mackenna
    4860, 7820436 Macul, Santiago, Chile}
\altaffiltext{2}{Departmento de Estad\'istica, Facultad de
  Matem\'aticas, Pontificia Universidad
  Cat\'olica de Chile, Av.\ Vicu\~na Mackenna
    4860, 7820436 Macul, Santiago, Chile}
\altaffiltext{3}{School of Physics, University of Exeter, Stocker
  Road, Exeter EX4 4QL, UK}
\altaffiltext{4}{CASA, Department of Astrophysical and Planetary Sciences, University of Colorado, Boulder, CO 80309, USA}
\altaffiltext{5}{Division of Geological and Planetary Sciences,
  California Institute of Technology, MC 170-25 1200, E. California
  Blvd., Pasadena, CA 91125, USA}
\altaffiltext{6}{Department of Astrophysical Sciences,
    Princeton University, NJ 08544, USA}
\altaffiltext{7}{Department of Astronomy \& Astrophysics, University of California, Santa Cruz, CA, 95064, USA}
\altaffiltext{8}{Harvard-Smithsonian Center for Astrophysics, 60 Garden Street, Cambridge, MA 02138, USA}
\altaffiltext{9}{Astrophysics Group, Keele University, Staffordshire
  ST5 5BG, UK}
\altaffiltext{10}{Department of Physics, and Kavli Institute for
  Astrophysics and Space Research, Massachusetts Institute of
  Technology, Cambridge, MA, 02139, USA}
\altaffiltext{11}{Alfred P. Sloan Fellow}
\altaffiltext{12}{Packard Fellow}
\altaffiltext{13}{Fellow of the Swiss National Science Foundation}

\begin{abstract}
  We present a ground based optical transmission spectrum of the
  inflated sub-Jupiter mass planet WASP-6b. The spectrum was measured
  in twenty spectral channels from 480 nm to 860nm using a series of
  91 spectra over a complete transit event. The observations were
  carried out using multi-object differential spectrophotometry with
  the IMACS spectrograph on the Baade telescope at Las Campanas
  Observatory. We model systematic effects on the observed light
  curves using principal component analysis on the comparison stars,
  and allow for the presence of short and long memory correlation
  structure in our Monte Carlo Markov Chain analysis of the transit
  light curves for WASP-6. The measured transmission spectrum presents
  a general trend of decreasing apparent planetary size with wavelength and
  lacks evidence for broad spectral features of Na and K predicted by
  clear atmosphere models. The spectrum is consistent with that
  expected for scattering that is more efficient in the blue, as
  could be caused by hazes or condensates in the atmosphere of
  WASP-6b. WASP-6b therefore appears to be yet another massive
  exoplanet with evidence for a mostly featureless transmission
  spectrum, underscoring the importance that hazes and condensates can
  have in determining the transmission spectra of exoplanets.
\end{abstract}

\keywords{planetary systems, planets and satellites: atmospheres,
  techniques: spectroscopic}

\section{Introduction}

Due to their fortitious geometry transiting exoplanets allow the
determination of physical properties that are inaccesible or hard to
reach for non-transiting system. One of the most exciting
possibilities enabled by the transting geometry is to measure
atmospheric properties of exoplanets without the need to resolve them
from their parent star through the technique of transmission
spectroscopy. In this technique, the atmospheric opacity at the planet
terminator is probed by measuring the planetary size via transit light
curve observations at different wavelengths. The measurable quantity
is the planet-to-star radius ratio as a function of wavelength,
$(R_p/R_*) (\lambda) \equiv k(\lambda)$, and is termed the
transmission spectrum. The measurement of a transmission spectrum is a
challenging one, with one atmospheric scale height $H$ translating to
a signal of order $2 H k \approx 10^{-4}$ for hot Jupiters
\citep[e.g.,][]{Brown:2001:transspec}. The requirements on precision
favour exoplanets with large atmospheric scale heights, large values
of $k$ (e.g., systems transiting M dwarfs) and orbiting bright targets
due to the necessity of acquiring a large number of photons to reach
the needed precision.

The first successful measurement by transmission spectroscopy was the
detection with the {\em Hubble Space Telescope (HST}) of absorption by
Na I in the hot Jupiter HD~209458b \citep{Charbonneau:2002:hd209}. The
signature of Na was 2--3 times weaker than expected from clear
atmosphere models, providing the first indications that condensates
can play an important role in determining the opacity of their
atmospheres as seen in transmission \citep[e.g.,][and references
therein]{Fortney:2005:condensates}. Subsequent space based studies
have concentrated largely on the planets orbiting the stars HD~209458
and HD~189733 due to the fact that they are very bright stars, and
therefore allow the collection of large number of photons even with
the modest aperture of space based telescopes. A recent study of all
the transmission spectra available for HD~189733, spanning the range
from 0.32 to 24~$\mu$m, points to a spectrum dominated by Rayleigh
scattering over the visible and near infra-red range with the only
detected feature being a narrow resonance line of Na
\citep{Pont:2013:hd189}. For HD~209458, \citet{Deming:2013:hd209}
present new WFC3 data combined with previous STIS data
\citep{Sing:2008a:hd209}, resulting in a transmission spectrum
spanning the wavelength range 0.3 to 1.6 $\mu$m. They conclude that
the broad features of the spectrum are dominated by haze and/or dust
opacity. In both cases the spectra are different from those predicted
by clear atmosphere models that do not incorporate condensates.

In order to further our understanding of gas giant atmospheres it is
necessary to build a larger sample of systems with measured
transmission spectra. Hundreds of transiting exoplanets, mostly hot
gas giants, have been discovered by ground based surveys such as
HATNet \citep{Bakos:2004:hatnet}, WASP \citep{Pollacco:2006:wasp},
KELT \citep{Pepper:2007:kelt}, XO \citep{McCullough:2005:XO}, TRES
\citep{Alonso:2004:tres1} and HATSouth \citep{Bakos:2013a}, with
magnitudes within reach of the larger collecting areas afforded by
ground based telescopes but often too faint for {\em HST}\footnote{The
  {\em Kepler} mission \citep{Borucki:2010:kepler} has discovered
  thousands of transiting exoplanet candidates but the magnitude of
  the hosts are usually significantly fainter than the systems
  discovered by ground based surveys, making detection of their
  atmospheres more challenging.}. The ground based observations have
to contend with the atmosphere and instruments lacking the space-based
stability of {\em HST}, but despite these extra hurdles the pace of
ground-based transmission spectra studies is steadily
increasing. Following the ground-based detection of Na I in HD189733b
\citep{Redfield:2008:hd189} and confirmation of Na I in HD209358b
\citep{Snellen:2008:hd209}, Na I has been additionally reported from
the ground in WASP-17b \citep{Wood:2011:w17,Zhou:2012:w17} and XO-2b
\citep{Sing:2012:xo2:na}. K I has been detected in XO-2b
\citep{Sing:2011:xo2:k} and the highly eccentric exoplanet HD80606b
\citep{Colon:2012:hd806:k}. All of these studies have used high
resolution spectroscopy or narrow band photometry to specifically
target resonant lines of alkali elements. Recently, a detection of
H$\alpha$ has been reported from the ground for HD~189733b
\citep{jensen:2012:ha}, complementing previous space-based detection
of Ly$\alpha$ and atomic lines in the UV with {\em HST} for HD~189733b
and HD~209358b \citep{Vidal-Madjar:2003:hd209, vidal-madjar:2004:hd209,lecavelier:2010:hd189}.

Differential spectrophotometry using multi-object spectrographs offer
an attractive means to obtain transmission spectra given the
possibility of using comparison stars to account for the various
systematic effects that affect the spectral time series
obtained. Using such spectrograps, transmission spectra in the optical
have been obtained for GJ~1214b \citep[][610-1000 nm with
VLT/FORS]{Bean:2011:gj1214} and recently for WASP29-b \citep[][515-720
nm with Gemini/GMOS]{Gibson:2013:w29}, with both studies finding
featureless spectra.  In the near infrared \citet{Bean:2013:w19}
present a transmission spectrum in the range 1.25 to 2.35
$\mu$m for WASP-19b, using MMIRS on Magellan.  In this work we present
an optical transmission spectrum of another planet, WASP-6 b, an inflated sub Jupiter
mass ($0.504 M_J$) planet orbiting a $V=11.9$ G dwarf
\citep{Gillon:2009a}, in the in the range 471--863 nm. 

\section{Observations}
\label{sec:obs}

The transmission spectrum spectrum of WASP-6b was obtained performing
multi-object differential spectrophotometry with the Inamori-Magellan
Areal Camera \& Spectrograph \citep[IMACS,][]{Dressler:2011:imacs}
mounted on the 6.5m Baade telescope at Las Campanas Observatory.
A series of 91 spectra of WASP-6 and a set of comparison stars were
obtained during a transit of the hot Jupiter WASP-6b in October 03
2010 with the f/2 camera of IMACS, which provides an unvignetted
circular field of view of radius $r\approx 12$ arcmin. The large field
of view makes IMACS a very attractive instrument for multi-object
differential spectrophotometry as it allows to search for suitable
comparison stars that have as much as possible similar magnitude and
colors as the target star. The median cadence of our observations was
224 sec, and the exposure time was set to 140 sec, except for the
first eight exposures when we were tuning the exposure level and whose
exposure times were $\{30, 120, 150, 150, 150, 130, 130, 130\}$
sec. The count level of the brightest pixel in the spectrum of WASP-6
was $\approx$ 43000 ADU, i.e. $\approx$ 65\% of the saturation level.
In addition to WASP-6 we observed 10 comparison stars of comparable
magnitude, seven of which had the whole wavelength range of interest
($\approx 4700-8600\ \angstrom$) recorded in the CCD with enough
signal-to-noise. The seven comparison stars we used are listed
in Table~\ref{tab:comp}. The integrated counts over the wavelength range of
interest for the spectrum of WASP-6 was typically $\approx
3.6\times10^8$ electrons, giving a Poisson noise limit for the
white-light light curve of $\approx 0.06$ mmag. Each star was observed through
a $10\times 10$ arcsec$^2$ slit in order to avoid the adverse effects
of variable slit losses. We used the 300-l+17 grating as dispersing
element, which gave us a seeing-dependent resolution $\Delta \lambda$
which was $\approx$ $5\ \angstrom$ under $0.7$ arcsec seeing and a
dispersion of $1.34 \angstrom$ per pixel. In addition to the science
mask, we obtained HeNeAr arc lamps through a mask that had slits at
the same position as the science mask but with slit widths in the
spectral direction of $0.7$ arcsec. Observing such masks is necessary
in order to produce well defined lines that are then used to define
the wavelength solution.

\begin{deluxetable}{c}
\tabletypesize{\scriptsize}
\tablecaption{List of comparison stars \label{tab:comp}}
\tablewidth{0pt}
\tablehead{
\colhead{2MASS identifier}}
\startdata
2MASS-23124095-2243232\\
2MASS-23124836-2252099\\
2MASS-23124448-2253190\\
2MASS-23124428-2256403\\
2MASS-23114068-2248130\\
2MASS-23113937-2250334\\
2MASS-23114820-2256592\\
\enddata
\end{deluxetable}

\begin{figure}
\plotone{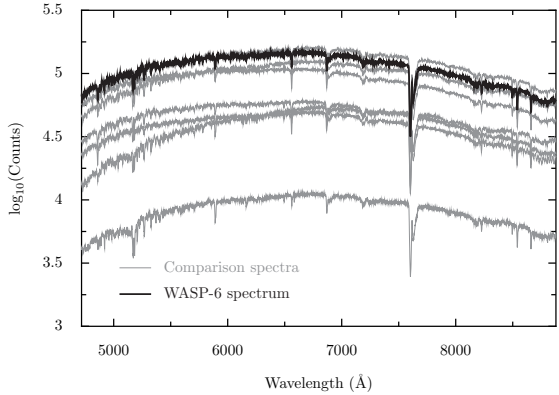}
\caption{Extracted spectra for WASP-6 and the seven comparison stars
  used in this work for a typical exposure.
\label{fig:W6Spec}}
\end{figure}

\begin{figure}
\plotone{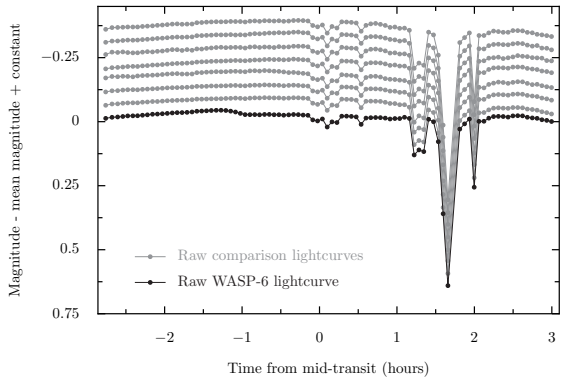}
\caption{Raw light curves for WASP-6 and seven comparison stars
  used in this work as a function of time.
\label{fig:counts}}
\end{figure}

The extracted spectra of WASP-6 and the seven comparison stars we used
are shown for a typical exposure in Figure~\ref{fig:W6Spec}. The
conditions throughout the night were variable. The raw light curves
constructed with the integrated counts over the whole spectral range
for WASP-6 and the comparison stars are shown as a function of time in
Figure~\ref{fig:counts}. Besides the variation due to varying airmass
(and the transit for WASP-6), there were periods with strongly varying
levels of transparency concentrated in the period of time 0--2 hrs
after mid-transit. The seeing was in the range $\approx$ 0.6--0.8\arcsec. 
In order  to maintain good sampling of the PSF in the spatial direction we 
defocused the telescope slightly in the periods of best seeing. Changes
in seeing and transparency left no noticeable traces in the final
light curves.

\section{Data reduction}

\subsection{Background and sky substraction}

After substracting the median value of the overscan region to every
image, an initial trace of each spectrum was obtained by
calculating the centroid of each row, which are perpendicular to the
dispersion direction. Each row was then divided in
three regions: a central region, which contains the bulk of the light
of the star, a middle, on-slit region which is dominated by sky continuum and
line emission, and an outer, out-of-slit region which contains a
smooth background outside the slit arising from, e.g., scattered
light. The middle and outer regions have components on each side of the spectrum.
The outermost region was used to determine a smooth background that
varies slowly along the dispersion direction. The median level was
obtained in the outer regions on either side of the slit, and then a
3rd order polynomial was used to estimate the average background level
as a function of pixel in the dispersion direction. This smooth
background component was then subtracted from the central and middle
regions. Then a Moffat function plus a constant level $c_i$ was fit
robustly to each background subtracted row in those regions. The
estimated $c_i$ (one per row), was then substracted from the central and
middle regions in order to obtain a spectrum where only the stellar
contribution remains. It is necessary to estimate the sky emission on
a row-per-row basis as sky emission lines have a wide, box-shaped
form with sharp boundaries due to the fact they fully illuminate the
wide slit.

\subsection{Fine tracing and spectrum extraction}

The background and sky subtracted spectrum was traced by an algorithm
that cross-correlates each slice perpendicular to the wavelength
direction with a Gaussian in order to find the spectral trace. The
centers of the trace were then fitted robustly with a fourth order
polynomial. This new tracing procedure served as a double check for
the centers obtained via the centroid method in the background and sky
substraction part of the data reduction process; both methods gave
traces consistent whith each other.  With the trace in hand, the
spectrum was extracted by using a simple extraction procedure,
i.e. summing the flux on each row $\pm 15$ pixels form the trace
position at that row. We also tried optimal extraction
\citep{Marsh:1989a}, but it lead to additional systematic effects when
analyzing the light curves,\footnote{Optimal extraction assumes 
that the profile along the wavelength direction is smooth enough 
     to be approximated by a low order polynomial. However, this assumption 
     is not always valid. In particular, we found that fringing in the reddest 
     part of the spectra induces fluctuations in the extracted flux with 
     wavelength due to the inadequacy of the smoothness assumption.} and in any case optimal extraction is not
expected to give significant gains over simple extraction at the high
signal-to-noise levels we are working with here. 
We also took
spectroscopic flats at the begining of the night with a quartz lamp
and reduced the data using both flat-fielded and non flat-fielded
spectra. The results where consistent when using both alternate
reductions, but the flat-fielded spectra showed higher dispersion in
the final transmission spectrum. We therefore used the non
flat-fielded spectra in the present work.

\subsection{Wavelength calibration}

The extracted spectra were calibrated using NeHeAr lamps taken at the
start of the night. The wavelength solution was obtained by the
following iterative procedure: pixel centers of lines with known
wavelengths where obtained by fitting Gaussians to them, and then all
the pixel centers, along with the known wavelengths of the lines were
fitted by a 6th order Chebyshev polynomial. We checked the
absolute deviation of each line from the fit, and removed the most
deviant one from our sample, repeating the fit without it.  This
process was iterated, removing one line at a time, until a rms of less
than 2000 m/s was obtained.  The rms of the final wavelength
solution was $\approx 1200$ m/s, using $27$ lines.

The procedure explained in the preceding paragraph served to
wavelength calibrate the first spectrum of the night closest in time
to the NeHeAr lamps. In order to measure and correct for wavelength
shifts throughout the night, the first spectrum was
cross-correlated with the subsequent ones in pixel-space in order to find the
shifts in wavelength-space. If $\lambda_{t_0,s}(p)$ is the wavelength
solution at time $t_0$ (the beggining of the night) for star $s$ as a
function of the pixel $p$, then the wavelength solution at time $t$ is
just $\lambda_{t,s}(p+\delta p_{t,s})$, where $\delta p_{t,s}$ is the
shift in pixel-space found by cross-correlating the spectrum of star
$s$ taken at time $t_0$ with the one taken at time $t$.
Finally, each spectrum was fitted with a b-spline in order to
interpolate each of the spectra into a common wavelength grid with
pixel size $0.75 \angstrom$.

\section{Modelling Framework}

The observed signal of WASP-6 is perturbed with respect to its
intrinsic shape, which we assume ideally to be a constant flux,
$F$. This constant flux is multiplied by the transit signal, $f(t;
\theta)$, which we describe parametrically using the formalism of
\citet{Mandel:2002a}. In what follows $\theta$ represents the vector
of transit parameters. The largest departure from this idealized model
in our observations will be given by systematic effects arising from atmospheric and
instrumental effects, which are assumed to act multiplicatively on our
signals. We will model the logarithm of the observed flux, $L(t)$, as

\begin{equation}
\label{eq1}
L(t; \theta) = S(t) + \log_{{10}}f(t; \theta)+\log_{{10}}F + \epsilon(t),
\end{equation}

\noindent where $S(t)$ represents the (multiplicative) perturbation to
the star's flux, which we'll refer to in what follows as the
perturbation signal, and $\epsilon(t)$ is a stochastic signal which
represents the noise in our measurements (under the term noise we will
also include potential variations of the star  that are not accounted
for in the estimate of the deterministic $S(t)$ and that can be
modelled by a stochastic signal).

\subsection{Modelling the perturbation signal}

\subsubsection{Estimation of Systematic Effects via Principal
  Component Analysis of the Comparison Stars}

Each star in the field is affected by a different perturbation
signal. However, these perturbation signals have in common that they arise
from the same physical and instrumental sources. In terms of
information, this is something we want to take advantage of. We model
this by assuming that a given perturbation signal is in fact a linear
combination of a set of signals $s_i(t)$, which represent the
different instrumental and atmospheric effects affecting all of our
lightcurves, i.e.,

\begin{equation}
\label{eq:pert}
S_k(t) = \sum_{i=1}^K \alpha_{k,i} s_i(t).
\end{equation}

Note that this model for the perturbation signal so far includes the
popular linear and polynomial trends (e.g., $s_i(t) = t^i$).
According to this model, the logarithm of the flux of each of $N$ stars
without a transiting planet in our field can be modelled as

\begin{equation}
L_{k}(t; \alpha) = S_{k}(t; \alpha) + \log_{{10}}F_{k} + \epsilon_{k}(t),
\label{eq:model_nt}
\end{equation}

\noindent where $\alpha$ denotes the set of parameters
$\{\alpha_{k,i}\}_{i=1}^K$. In the case in which we have a set of
comparison stars, we can see each of them as an independent (noisy)
measurement of a linear combination of the signals $s_i(t)$ in
Eq.~(\ref{eq:pert}). A way of obtaining those signals is by assuming
that the $s_i(t)$ are uncorrelated random variables, in which case
these signals are easily estimated by performing a principal component
analysis (PCA) of the mean-substracted lightcurves of the comparison
stars. Given $N$ comparison stars one can estimate at most $N$
components, and thus we must have $K \leq N$. As written in
Eq.~\ref{eq:model_nt} we cannot separate $s_i(t)$ from
$\epsilon_k(t)$, and in general the principal components will have
contributions from both terms. If $s_i(t) \gg \epsilon_k(t)$ the $K$
principal components that contribute most to the signal variance will
be dominated by the perturbation signals, but some projection of the
$\epsilon_k$ into the estimates of $s_i$ is to be expected.

\subsubsection{Selecting the number of principal components}

In our case, the number of components $K$ is unknown a priori. We need
therefore to determine an optimal number of principal components
to describe the perturbation signal, taking in consideration that
there is noise present in the lightcurves of the comparison stars and,
thus, some of the principal components obtained are mostly
noise. There are several possibilities for doing this depending on
what we define as optimal. We will determine the optimal number
of components as the minimum number of components that are able to
achieve the best predictive power allowed by the maximum set of $N$
components available.

As a measure of predictive power we use $k$-fold cross-validation
procedure \citep{ESL07}. $k$-fold cross-validation is a procedure
which estimates prediction error, i.e., how well a model predicts
out-of-sample data. The idea is to split the datapoints in $k$
disjoint groups (called folds).  A``validation'' fold is left out and
a fit is done with the remaining ``training'' folds, allowing to
predict the data in the validation fold that was not used in this
fitting procedure. This procedure is repeated for all folds. Denoting
the datapoints by $y_i$ and the values predicted on the $k$-th
fold by the cross-validation procedure by $f_i^{-k}$, an estimate for
the prediction error is

\begin{equation}
\label{cv_error}
\hat{CV} = \frac{1}{N} \sum_{i=1}^{N}\mathcal{L}(y_i^k-f^{-k}_i), \nonumber
\end{equation}

\noindent where $\mathcal{L}(\cdot)$ is the loss function. Examples of
loss functions are the $\mathcal{L}_1$ norm ($\mathcal{L}_1(x) = |x|$)
or the $\mathcal{L}_2$ norm ($\mathcal{L}_2(x) = x^2$).

In our case, the light curves of the $N$ comparison stars are used to
estimate $l<N$ principal components. These $l$ principal components,
which are a set of light curves $\{s_i\}_{i=1}^l$, are our estimates
of the systematic effects, and we use the out-of-transit part of the light
curve of WASP-6 as the validation data by fitting it with the
$\{s_i\}_{i=1}^l$. In more precise terms, if $y(t_k)$ denote the time
series of the out-of-transit portion of the light curve of WASP-6, we
apply $k$-fold cross-validation by considering a model of the form
$y(t_k) = \sum_{i=1}^l \alpha_is_i(t_k)$.

\subsection{Joint Parameter Estimation for Transit and Stochastic Components}

\begin{figure}
\plotone{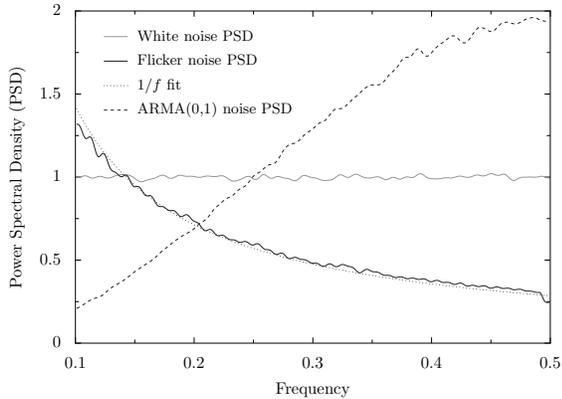}
\caption{Example of the structure that is expected in the power
  spectral density (PSD) of
  residual signals of the different types considered in this paper. 
The PSDs shown here are the mean of $10000$ realizations 
with the noise-structures indicated in the figure. Note that the 
white-noise PSD is flat, while the flicker-noise and the ARMA($0,1$) models cover low and high-frequency ranges. 
respectively.
\label{noise_graph}}
\end{figure}

In the past sub-sections we set up an estimation process for the
signal given in eq. (\ref{eq:pert}) using principal component
analysis. It remains to specify a model for the stochastic signal
which we have termed noise, i.e., the $\epsilon(t)$ term in
eq. (\ref{eq1}). As noted above, the principal components will absorb
part of the $\epsilon(t)$, and so our estimate of the noise may not
necessarily accurately reflect the $\epsilon(t)$ term in
eq. (\ref{eq1}) assuming the model holds. Nonetheless, this is of no
consequence as we just aim to model the residuals after the time
series has been modeled with the $\{s_i\}_{i=1}^l$. While we still
call this term $\epsilon(t)$ in what follows, one should bear in mind
this subtlety. 
%
%
An important feature of the correlated stochastic models we consider
is that they can model trends.
%
%
The $\{s_i\}_{i=1}^l$ are obtained from the comparison
stars, and while the hope is that they capture all of the systematic
effects, it is possible that some systematic effects individual to the
target star are not captured. The stochastic ``noise'' models
considered below that have time correlations can in principle capture
remnant individual trends particular to WASP-6. 

We make use of Markov Chain Monte Carlo \citep[MCMC; see,
e.g.,][]{Ford:2005a} algorithms to obtain estimates of the posterior
probabilty distributions of our parameters, $\theta, \alpha, \eta$,
given a dataset $\mathbf{y}$, where we have introduced a new set of
parameters $\eta$ characterizing a stochastic component (see below).
The posterior distribution $p(\mathbf{\theta, \alpha,
  \eta}|\mathbf{y})$ is obtained using a prior distribution for our
parameters $p(\mathbf{\theta, \alpha, \eta})$ and a likelihood
function, $p(\mathbf{y}|\mathbf{\theta, \alpha, \eta})$. Following
previous works \citep[e.g.][]{Carter:2009a,Gibson:2012:GP} we assume
that the likelihood function is a multivariate Gaussian distribution
given by

\begin{eqnarray}
\label{eq:lik}
p(\mathbf{y}|\mathbf{\theta, \alpha,\eta}) &=&
\frac{1}{(2\pi)^{n/2}|\mathbf{\Sigma_\eta}|^{1/2}} \\
 & & \times \exp\left[{-\frac{1}{2}
(\mathbf{y}-\mathbf{g}(\theta,
\alpha))^T\mathbf{\Sigma_\eta}^{-1}(\mathbf{y}-\mathbf{g}(\theta,
\alpha))}\right], \nonumber
\end{eqnarray}

\noindent where $\mathbf{g}(\theta, \alpha)$ is the function that predicts the
observed datapoints and $\mathbf{\Sigma}_\eta$ is the covariance
matrix which depends on the set of parameters $\eta$. It
is the structure of this matrix which defines the type of noise of the
residuals. 
Previous works have proposed to account for time correlated structure
in the residuals using flicker noise models, where it is assumed that
the noise follows a Power Spectral Density (PSD) of the form $1/f$
\citep{Carter:2009a}, and Gaussian processes, where the covariance
matrix is parametrized with a particular kernel that can incorporate
correlations depending on a set of input parameters, including
time \citep{Gibson:2012:GP,Gibson:2013:w29}
\footnote{In
  \citet{Gibson:2012:GP} a set of optical state parameters is used
  within a Gaussian process framework to model what we have termed
  here the perturbation signal, while in \citet{Gibson:2013:w29} the
  Gaussian process is used to account for the time correlation
  structure of the residuals in a procedure more comparable to ours.}.
In the present work we consider three different models: a white-noise
model, where the covariance matrix is assumed to be diagonal, a
flicker-noise model, and ARMA$(p,q)$ models, where the structure of
the covariance is determined via the parameters $p$ and $q$ (see
\S\ref{ssec:ARMA} below for the definition of ARMA$(p,q)$ models). The
reason for choosing these three models is that they sample a wide
range of spectral structure of the noise: white-noise models define
models where the PSD is flat, while flicker and ARMA-like models
define noise structures with PSDs with power in low and high
frequencies, respectively.  Figure~\ref{noise_graph} illustrates the
various structures the PSD can have for the different noise structures
considered here.

\subsubsection{Flicker noise model}

Flicker-noise is known to arise in many astrophysical time series
\citep{Press:1978a}.  It is a type of noise that fits long-range
correlations in a stochastic process very well because of its assumed
PSD shape of $1/f$. An efficient set of algorithms for its
implementation in MCMC algorithms was proposed recently by
\citet{Carter:2009a}. The basic idea of this implementation is to
assume that the noise is made up of two components: an uncorrelated
Gaussian process of constant variance and a correlated Gaussian
process which follows this flicker-noise model. These two components
are parametrized by $\sigma_w$ and $\sigma_r$,
characterizing the white and correlated noise components,
respectively. A wavelet transform of the residuals takes the problem
into the wavelet basis where flicker noise is nearly-uncorrelated
making the problem analytically and computationally more tractable.

\subsubsection{ARMA noise model}
\label{ssec:ARMA}

Autoregressive-moving-average (ARMA) models have been in use in the
statistical literature for a long time with a very broad range of
different applications \citep{TSA91}. Although known for long in the
astronomical community \citep[e.g.,][]{Scargle:1981a,Koen:1993a}, these
noise models haven't been used so far for transit lightcurves to the
best of our knowledge.\footnote{ARMA$(p,q)$ models have been considered
  recently in the modeling of radial velocity data
  \citep{Tuomi:2013:tceti}. The very iregular sampling
in those data needs careful consideration, in the case of transit
light curve analysis their use is more direct given the nearly uniform
sampling that is obtained for these observations.}  

The time series $X_{t_k}$ of an ARMA($p,q$) process, where the $t_k$
are the times of each observation, satisfies

\begin{equation}
X_{t_k}= \sum_{i=1}^p \phi_i X_{t_{k-i}}+\sum_{i=1}^q \theta_i \varepsilon(t_{k-i})+\varepsilon(t_k).
\end{equation}

\noindent where the $\{\phi_i\}^p_{i=1}$ and $\{\theta_i\}^q_{i=1}$
are the parameters of the model and $\varepsilon_{t_k}$ is white noise
with variance $\sigma_w^2$. The orders $(p,q)$ of the
ARMA($p,q$) model define how far in the past a given process looks
at when defining future values. Long-range correlations need a high
order ARMA model, while short-range correlations need lower order
models. An ARMA model allows us to explore a higher range of noise
structures in a complementary way to flicker noise models.

In order to fit ARMA models to the residuals via an MCMC algorithm, we
need the likelihood function of the model given that the residuals
follow an ARMA($p,q$) model. For this we implented the recursive
algorithm described in \citet[Chapter 8]{TSA91}, which assumes that
$\varepsilon (t_k)$ follows a normal distribution with constant
variance and that the ARMA process is causal and invertible.

\subsubsection{Stochastic model selection}
\label{ssec:model_sel}

Given the three proposed noise models for the stochastic signal
$\epsilon(t)$, it remains to define which of the three affords a
better description of the data, taking into account the trade-off between
the complexity of the proposed model and its goodness-of-fit.
There are several criteria for model selection, a comprehensive
comparison between different criteria has been done recently by
\citet{Vehtari:2012a}. The main conclusion is that, despite the fact that many
model selection criteria have good asymptotic behavior under the
constraints that are explicit when deriving them, there is no
``perfect model selection'' criteria, and there is a need to compare
the different methods in the finite-sample case. Following this
philosophy, we compare in this work the results of the AIC \citep[``An
Information Criterion'';][]{Akaike:1974a}, the BIC \citep[``Bayesian
Information Criterion'';][]{Schwarz:1978a}, the DIC \citep[``Deviance
Information Criterion'';][]{Spiegelhalter:2002a} and the DIC$^{\rm{A}}$,
a modified version of DIC with a proposal for bias-correction \citep{Ando:2012a}.

\section{Lightcurve analysis}

From the initial ten comparison stars, only seven were used to
correct for systematic effects. One star was eliminated on the grounds
of having significantly less flux than the rest and the other two due
to not having the whole spectral range of interest recorded in the
CCD.  Given the seven comparison stars, we applied PCA to the
mean-substracted time-series in order to obtain an estimate of the
perturbation signals. We describe now the construction and analysis of
the white light transit light curve and the light curves for 
20 wavelength bins.

\subsection{White light transit light curve}
\label{ssec:wl}

\begin{figure}
\plotone{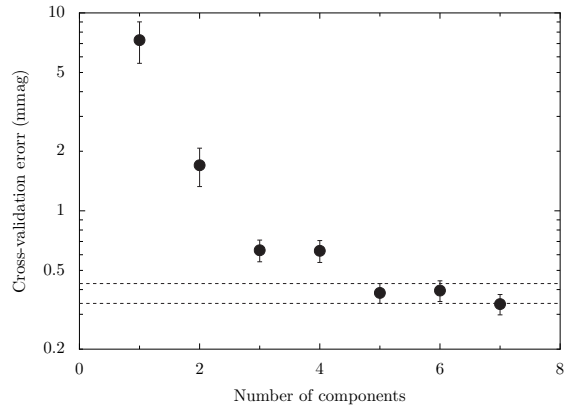}
\caption{Cross-validation error for the prediction of out-of-transit data using different number of principal 
components with the $5$-fold cross-validation procedure we adopted. Note that the minimum is at $k=7$ (dashed lines 
indicate the value at the minimum and a value higher by $1\sigma$), but the value at
$k=5$ achieves similar error with lower degrees of freedom.
\label{cv_graph_wl}}
\end{figure}

In order to obtain the white light transit light curve of WASP-6, we
summed the signal over the wavelength range $4718$ to $8879$ \AA\ for
the target and the comparison stars. Then, we performed $5$-fold
cross-validation in the out-of-transit part of the light curve of
WASP-6 in order to obtain the optimal number of components to be used in
our MCMC algorithm. The result of this cross-validation procedure is
shown in Figure \ref{cv_graph_wl}.

From the results of our $5$-fold cross-validation procedure one may
choose either $k=7$ (the value of the minimum error) or $k=5$, which
is at less than 1-standard error away from the value at the minimum.
We choose this last value because it allows a similar prediction error
as the minimum with two less parameter.

\begin{figure}
\plotone{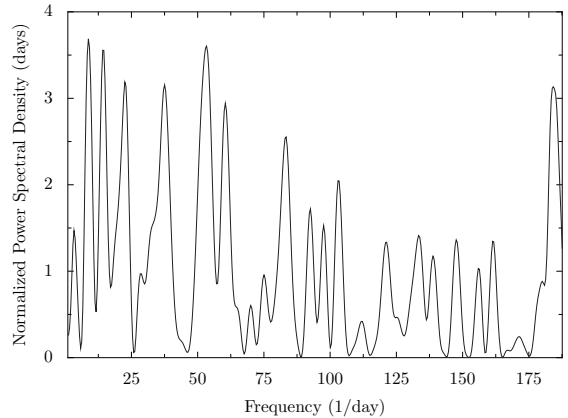}
\caption{Power spectral density of the residuals of the fit using
  white Gaussian noise (see Figure~\ref{white_light_tl} to see the
  residuals). Note the preference for high power at small frequencies.
  \label{res_PSD}}
\end{figure}

Using the first five principal components, we fitted the model
proposed in Equation~\ref{eq1} first using a white Gaussian noise model
via MCMC using the PyMC python module \citep{Patil:2010a}. We used wide
truncated Gaussian priors\footnote{We denote our truncated Gaussian
  priors as TruncNorm$(\mu,\sigma^2)$. They are Normal
  distributions restricted to take values in the range ($0,\infty$), i.e. they are
  restricted to be positive.} in order to incorporate previous
measurements of the transit parameters obtained by \citet{Gillon:2009a} and
\citet{Dragomir:2011a}, and the orbital parameters of \citet{Gillon:2009a}. We
adopt a quadratic limb darkening law of the form $I(\mu) =
I(1)[1-u_1(1-\mu)-u_2(1-\mu)^2]$, where $\mu=\cos(\theta)$ and
$\{u_1,u_2\}$ are the limb darkening coefficients. It is well known
that $u_1$ and $u_2$ are strongly correlated \citep{Pal:2008a}, and it has been shown
that if we define new coefficients $(w_1,w_2)$ that are related to
$(u_1,u_2)$ by $(w_1,w_2) =R(\pi/4)(u_1,u_2)$ where 

\[
R(\theta) = \left( 
\begin{array}{cc} 
\cos(\theta) &  -\sin(\theta)\\
\sin(\theta) & \cos(\theta)\\ 
\end{array} \right)
\] 

\noindent is a rotation matrix by $\theta$ radians, then $w_1, w_2$
are nearly uncorrelated and transits are mostly sensitive to $w_1$,
with $w_2$ essentially constant \citep{Howarth:2011a}. In our MCMC
analysis we fix $w_2$ to the (wavelength dependent) value calculated
for the stellar parameters of WASP-6 as described in
\citet{Sing:2010a}.

\begin{figure*}
\plotone{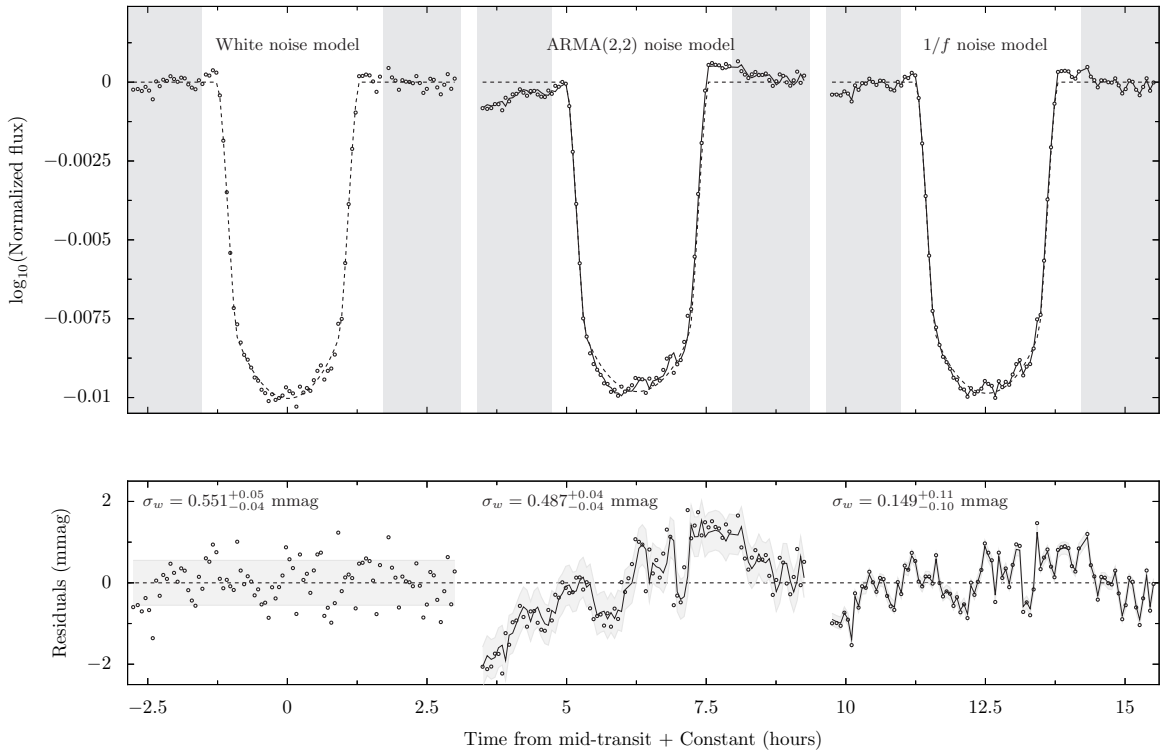}
\caption{({\em Top}) The circles show the baseline-substracted lightcurves (i.e.,
  lightcurves with the fitted perturbation signal substracted) using
  the  different noise models indicated. We also show the
  corresponding best-fit transit models (dashed line) and the best-fit
  transit models plus an estimate of the correlated noise component (solid
  line, only for the two right-most lightcurves). The shaded regions indicate points that where
  used as out-of-transit data by the $5$-fold cross-validation
  procedure that selected the number of principal components to use in
  the fits.
 ({\em Bottom}) Residuals between the best-fit transit model and the
 baseline subtracted lightcurves (circles). The solid lines
in the two right-most set of points indicate estimates of the
correlated components obtained by projecting the residuals into the 
best-fit correlated component model (see \S5). The difference between
the points and the solid lines (dashed line for the white Gaussian
noise case) is the white Gaussian noise
component, whose dispersion  $\sigma_w$ is indicated for each of the
noise models considered and also illustrated with $\pm 1\sigma_w$
bands. 
\label{white_light_tl}}
\end{figure*}

Five MCMC chains of $10^6$ links each, plus $10^5$ used for burn-in,
where used. We checked that every chain converged to similar values
and then thinned the MCMC samples by $10^4$ in order to get rid of the
auto-correlation between the links. We used the thinned sample as our
posterior distribution, using the posterior median as an estimate of
each parameter (using the point in the chain with the largest
likelihood leads to statistically indistinguishable results). The fit using a white
Gaussian model for the noise allows us to investigate the structure of
the residuals, which show clear long-range correlations, as is evident
in the power spectral density of the residuals plotted in
Figure~\ref{res_PSD}. Note that the power is significantly higher at
lower frequencies, which suggest that the residuals have long-range
correlations. We performed an MCMC fit using an $1/f$-like and another
MCMC fit using an ARMA-like model for the residuals. Note that in
order to fit an ARMA($p$,$q$) model with our algorithms, we need to
define the order $p$ and $q$ of the ARMA process. In order to do this,
we fitted several ARMA($p$,$q$) models to the residuals of the white
Gaussian MCMC fit for different orders $p$ and $q$ using maximum
likelihood, and calculated the AIC and BIC of each fit. In the sense
of minimizing these information criteria, the ``best'' ARMA model was
an ARMA($2$,$2$) model, so we performed our MCMC algorithms assuming
this as the best model for the ARMA case. The results of the MCMC fits
assuming a white Gaussian noise model, an ARMA(2,2) noise model and a
$1/f$ noise model for the residuals are shown in
Figure~\ref{white_light_tl}, and a summary of the values of the
information-criteria for each of our MCMC fits are shown in
Table~\ref{tbl-1}.

\begin{deluxetable}{clll}
\tabletypesize{\scriptsize}
\tablecaption{Values for the different information criteria (IC) for each noise-model considered in our MCMC 
fits. \label{tbl-1}}
\tablewidth{0pt}
\tablehead{
\colhead{IC} & \colhead{WG model$^\textrm{a}$} & \colhead{ARMA$(2,2)$ model$^\textrm{a}$} & 
\colhead{$1/f$ model$^\textrm{a}$}
}
\startdata
AIC & -1260.59 & -1273.38 & -1833.72 \\
BIC & -1230.46 & -1234.20 & -1801.08 \\
DIC & -1165 & -1202.03 & -1793.60 \\
DIC$^\textrm{A}$ & -1105.20 & -1149.86 & -1760.53\\
\enddata
\tablenotetext{a}{Note that each of the noise models has a different number of 
parameters: the white Gaussian noise model (WG model) has 12 parameters, the ARMA($2$,$2$)-like 
noise model 16 parameters and the $1/f$-like noise model has 13 parameters.}
\end{deluxetable}

\begin{deluxetable*}{clll}
\tabletypesize{\scriptsize}
\tablecaption{MCMC priors used for the white light transit analysis. \label{tbl-3}}
\tablewidth{0pt}
\tablehead{
\colhead{Transit parameter} & \colhead{Description} & \colhead{Prior$^\textrm{a}$} & \colhead{Units}
}
\startdata
$R_p/R_s$ & Planet-to-star radius ratio & TruncNorm$(0.14,0.01^2)^A$ & -\\
$t_0$ & Time of mid-transit & TruncNorm$(55473.15,0.01^2)^c$ & MHJD\\
$P$ & Period & TruncNorm$(3.36,0.01^2)^b$ & days\\
$i$ & Inclination & TruncNorm$(1.546,0.017^2)^b$ & Radians\\
$R_s/a$ & Stellar radius to semi-major axis ratio & TruncNorm$(0.09,0.01^2)^b$ & - \\
$w_1$ & Linear limb-darkening coefficient & U$(0,1)$ & - \\
$\sigma_w$ & Standard deviation of the white noise part of the noise model & U$(0,1)$  & mag\\
$\sigma_r$ & Noise parameter for the $1/f$ part of the noise model$^\textrm{d}$ & U$(0,1)$ & mag\\
\enddata
\tablenotetext{a}{The TruncNorm$(\mu,\sigma^2)$ distributions are Normal
  distributions truncated to take values in the range ($0,\infty$),
  i.e., they are required to be positive. The U$(a,b)$ distributions are uniform distributions
  between $a$ and $b$.}  
\tablenotetext{b}{Obtained from the
  arithmetic mean between the values cited in \cite{Gillon:2009a} and
  \cite{Dragomir:2011a}. The variance of the prior covers more than
  $3\sigma$ around their values.}
\tablenotetext{c}{Obtained from the values cited in
  \cite{Gillon:2009a}. The variance of the prior covers more than
  $3\sigma$ of their values.}
\tablenotetext{d}{Not to be interpreted as the standard-deviation of the $1/f$ part of the noise.}
\end{deluxetable*}

\begin{deluxetable*}{clll}
\tabletypesize{\scriptsize}
\tablecaption{WASP-6b transit parameters estimated using the white light transit
 lightcurve using a $1/f$-like noise model. \label{tbl-2}}
\tablewidth{0pt}
\tablehead{
\colhead{Transit parameter} & \colhead{Description} & \colhead{Posterior value} & \colhead{Units}
}
\startdata
$R_p/R_s$ & Planet-to-star radius ratio & $0.1404^{+0.0010}_{-0.0010}$ & -\\
$t_0$ & Time of mid-transit & $55473.15365^{+0.00016}_{-0.00016}$ & MHJD\\
$P$ & Period & $3.3605^{+0.0098}_{-0.0101}$ & days\\
$i$ & Inclination & $1.5465^{+0.0074}_{-0.0055}$ & Radians\\
$R_s/a$ & Stellar radius to semi-major axis ratio & $0.0932^{+0.0015}_{-0.0015}$ & - \\
$w_1$ &  limb-darkening coefficient (see \S~\ref{ssec:wl}) & $0.44^{+0.12}_{-0.12}$ & - \\
$\sigma_w$ & Standard deviation of the white noise part of the noise model & $0.1492^{+0.1078}_{-0.1021}$  & mmag\\
$\sigma_r$ & Noise parameter for the $1/f$ part of the noise model$^\textrm{a}$ & $3.26^{+0.03}_{-0.50}$ & mmag\\
\enddata
\tablenotetext{a}{Not to be interpreted as the standard-deviation of
  the $1/f$ part of the noise.}
\label{tab:wl_transit}
\end{deluxetable*}

It is important to note that the residuals shown in
Figure~\ref{white_light_tl} are the signal left over after subtracting
the deterministic part of the model only (denoted by
$\mathbf{g}(\theta, \alpha)$ in equation~\ref{eq:lik}). Therefore,
they still contain in the case of the ARMA(2,2) and $1/f$ noise
models, a correlated stochastic component summed with a white noise
component. As opposed to deterministic components, the stochastic
components cannot just be predicted given the times $t_i$ of the
observations, as we only know the distribution of expected values once
we know the parameters ($\{\theta_1,\theta_2,\phi_1,\phi_2,\sigma_w\}$
for ARMA(2,2), $\{\sigma_r,\sigma_w\}$ for flicker noise and
$\sigma_w$ for white Gaussian noise).
But even though we cannot plot a unique expected trend given the
best-fit parameters for the correlated noise models, we can apply
filters to the residuals that project them into the best-fit model, or
viewed differently, we can filter out the expected white Gaussian
noise component leaving just the correlated
part. Such filters allow us then to build {\em estimates} of the
particular realization of a given process that is present in our
residuals. For the ARMA(2,2) and $1/f$ case we plot in the bottom
panel of Figure~\ref{white_light_tl} estimates of the correlated
components as solid lines through the residuals\footnote{For the $1/f$
  model we use the whitening filter presented in \citet[][see
  \S3.4]{Carter:2009a}, while for the ARMA(2,2) process we use
  prediction equations in the time domain \citep[][see
  \S5.1]{TSA91}.}. It is the difference between these lines and the
residual points that constitute the remaining white Gaussian noise
component with dispersion $\sigma_w$ indicated in the residuals panel.

It is informative to discuss the different values of the $\sigma_w$
parameter inferred for each of the models we consider. For the
white Gaussian noise model, the value of this parameter is
$\sigma_w=0.55^{+0.05}_{-0.04}$ mmag, which is an order of magnitude
higher than the underlying Poisson noise ($\approx$ 0.06 mmag,
see \S~\ref{sec:obs}). The same goes for the ARMA-like noise model fit,
which has a value of this parameter of
$\sigma_w=0.49^{+0.04}_{-0.04}$ mmag. Finally, for the $1/f$-like noise
model the value of this parameter is $\sigma_w=0.15^{+0.11}_{-0.10}$ mmag,
which is just $\approx 2.5$ times the Poisson noise limit. Motivated
by this result and by the values of the information-theoretic model
selection measures quoted on Table~\ref{tbl-1}, we conclude that the
preferred model is the one that models the underlying stochastic
signal as $1/f$-like noise. We note that as \citet{Carter:2009a}
stress in their work, using this model for the residuals increases the
uncertainty in the transit parameter, but provides more realistic
estimates for them. We select the model parameters fitted using the
$1/f$ noise model, which are quoted in Table~\ref{tbl-2}, as the
best estimates from now on. These parameters are generally an
improvement on previous measurements by \citet{Gillon:2009a} and
\citet{Dragomir:2011a}.

We close this section by noting that the principal component
regression we adopted was able to recover from the high periods of
absorption present 0--2 hrs after mid-transit (see
Figure~\ref{fig:counts}) without leaving a noticeable trace in the final
light curve.

\subsection{WASP-6b transmission spectrum}

\begin{figure*}
\plottwo{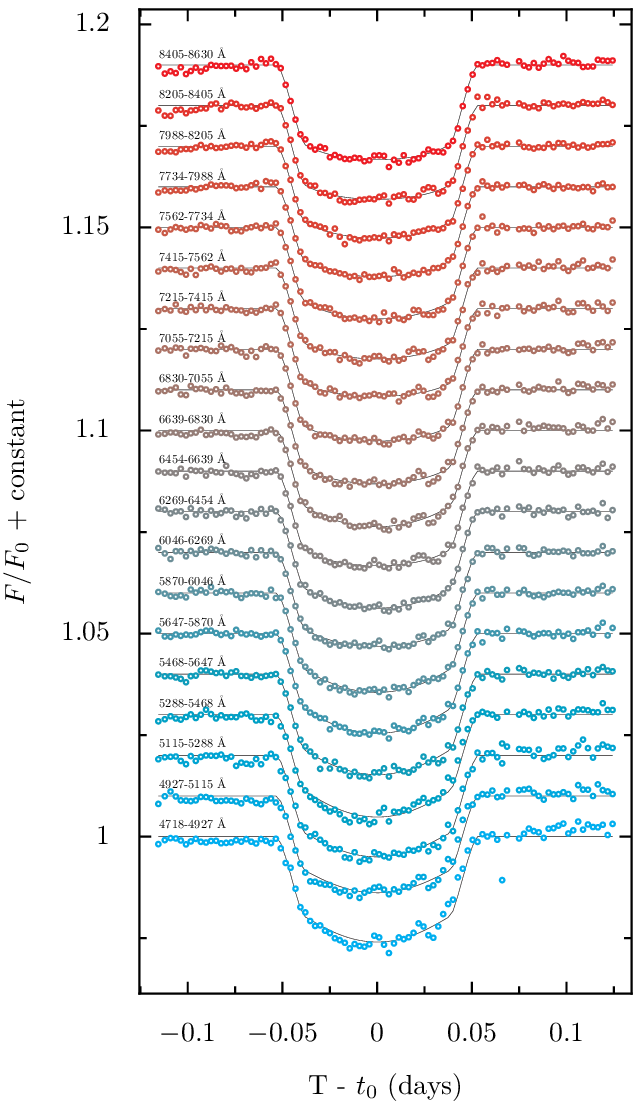}{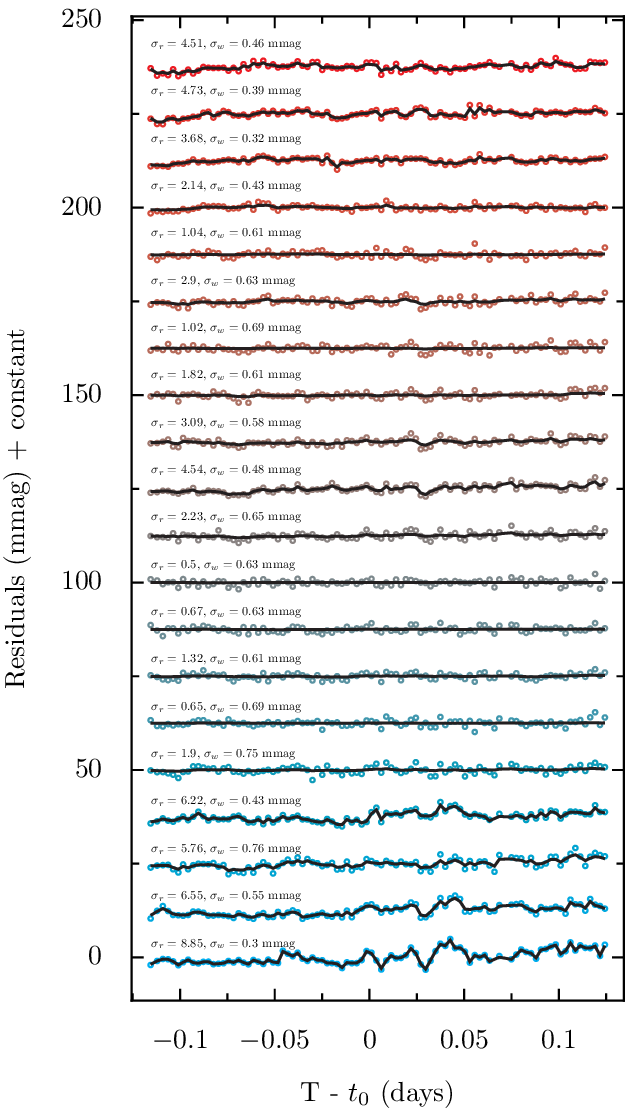}
\caption{ ({\em Left}) Transits as observed in different wavelength channels along with the best-fit transit signal plus 
the stochastic $1/f$ noise signal. The obvious outlier close to $t-t_0 =0.05$ at the first, bluest, 
wavelength bin was not included in the fit. ({\em Right}) Residuals between the best-fit transit model and the
 baseline subtracted lightcurves for each of the wavelength channels (circles). The solid lines
indicate estimates of the
correlated $1/f$ component obtained as described in \S5. The best fit
parameters of the $1/f$ component are indicated over each set of residuals.
\label{fig:transits}}
\end{figure*}

\begin{figure*}
\plotone{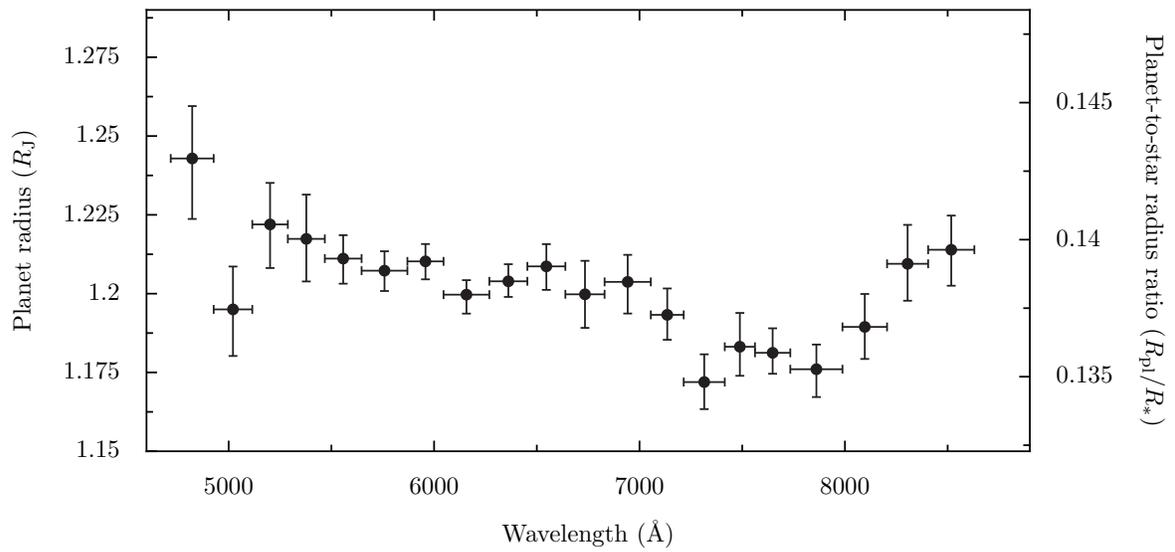}
\caption{Transmission spectrum of WASP-6b measured with IMACS.
\label{fig:transpec_dataonly}}
\end{figure*}

The procedures explained in the previous sub-section were replicated
for the time series of each of 20 wavelength bins, but now leaving
only the planet-to-star radius ratio, the linear limb darkening
coefficient $w_1$ and the noise parameters as free parameters (all the
other transit parameters where fixed to the values shown in Table
\ref{tbl-2} while the values for $w_2$ are calculated as indicated in
\S~\ref{ssec:wl} and are indicated in Table~\ref{tab:transits}). Priors were the same as the white light analysis
for parameters for $\mu_1, \sigma_r$ and $\sigma_w$, and the MCMC
chains were set-up similarly except that a thinning value of $10^3$
was used. The prior for $(R_{pl}/R_*)$ was set to
$TruncNorm(0.1404,0.01^2)$, i.e.\ we set the mean to the posterior
value of our white light analysis\footnote{}. The wavelength bins were chosen to
be $\approx$ 200 \AA\ wide, with boundaries that lie in the
pseudo-continuum of the WASP-6 spectrum, as boundaries in steep parts
of the spectrum such as spectral lines would in principle maximize
redistribution of flux between adjacent bins under the changing seeing
conditions which set the spectral resolution in our setup. For a given
spectral bin, the number of principal components was selected
separately because different systematics may be dependent on
wavelength, and therefore the number of principal components needed
may change. In practice, no more than one principal component was
added or subtracted in each wavelength bin when compared to the five
components used for the white light curve. In all of them, however,
the noise model to be used is the same, the $1/f$-like noise
model. Figure~\ref{fig:transits} shows the baseline-substracted data
along with the best-fit transit model at different wavelengths, and
Table~\ref{tab:transits} tabulates the transit parameters from the
MCMC analysis for each wavelength bin. The values of $R_{\rm pl}/R_*$
as a function of wavelength constitute our measured transmission
spectrum which is shown in Figure~\ref{fig:transpec_dataonly}; the
typical uncertainty in $R_{\rm pl}/R_*$ is $\approx 0.8$\%, and the
inferred $\sigma_w$ values are typically $\approx$ 1--3 times the
Poisson limit in each wavelength bin, for which a typical value is
0.25 mmag.

\subsection{Limits on the contribution of unocculted stellar spots}

As pointed out in several works
\citep[e.g.,][]{Pont:2008:hd189,Sing:2011:hd189}, stellar spots --
both occulted and unocculted during transit -- can affect the
transmission spectrum. In our transit light curve we see no
significant deviations that could be attributed to an occulted
starspot, so in what follows we estimate the potential signal induced
in the transmission spectrum by unocculted stellar spots.

Stellar spots can be modelled as regions in the surface of the star
that have a lower effective temperature that the photosphere. Given
that WASP-6 is a G star, we can use the Sun as a proxy to infer spot
properties. Sunspots can be characterized as having a temperature
difference with the photosphere of $\Delta T \approx -500$ K
\citep[][see \S2.2]{Lagrange:2010:spots}. This is an effective value
that represents a good average for the different
values of $\Delta T$ in the umbral and penumbral regions. Given a
fraction of the stellar surface $f_s$ covered by spots characterized
by temperature $T+\Delta T$, the total
brightness of the star will be changed by a factor $1 + f(\lambda) = 1
+ f_s (I_\lambda(T+\Delta T, \theta) / I_\lambda(T,\theta) - 1)$, where
$I_\lambda(T, \theta)$ is the surface brightness of a star with
effective temperature $T$ and other stellar parameters given by
$\theta = (\log g, Z,\ldots)$. If the fractional change in flux $\epsilon$ caused by
spots at a reference wavelength $\lambda_0$ can be measured then $f_s$
can be inferred to be $f_s = \epsilon / ( I_{\lambda_0}(T+\Delta T,\theta)/I_{\lambda_0}(T,\theta) -
1)$ and then we can
write $f(\lambda) = \epsilon (I_\lambda(T+\Delta T, \theta) / I_\lambda(T,\theta) - 1)
(I_{\lambda_0}(T+\Delta T, \theta) / I_{\lambda_0}(T,\theta) - 1)^{-1}$ 
\citep[c.f.][eq. 4]{Sing:2011:hd189}.

A change in the stellar luminosity due to starspots will have an effect
on the measured value of $k = R_p/R_*$, and as the effect is
chromatic, it will induce an effect in the tranmission spectrum. The
decrease of flux during transit with respect to the out of transit
flux $F_0$ is given by $ (\Delta F/ F_0) = k^2$ (neglecting any
emission from the planet). If $F_0$ is changed by starspots by a
fractional amount $f(\lambda)$ we have $\delta (\Delta F / F_0)
\approx -(\Delta F / F_0) \delta F_0 / F_0 \equiv -(\Delta F / F_0)
f(\lambda) = k^2 f(\lambda) = 2k \delta k$, where we have used
$f(\lambda) \equiv \delta F_0 / F_0$. From here we get
finally\footnote{This is a special case of the derivation of
  \citet{Desert:2011:hd189}, namely their case $\alpha=-1$ which
  corresponds to neglecting changes in brightness of the fraction of
 the  stellar disk that is not affected by spots.}

\[
\frac{\delta k}{k} = \frac{-f(\lambda)}{2}.
\]

\begin{figure}
\plotone{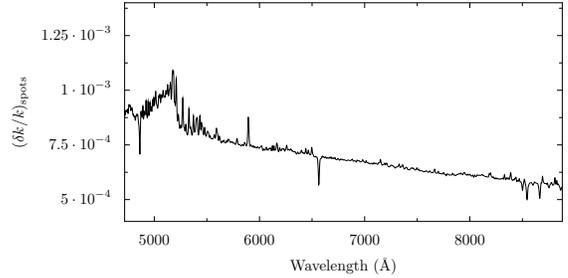}
\caption{Predicted fractional change in $k=(R_p/R_*)$ due to stellar
  spots that produce a rotation amplitude of $\epsilon=-4$ mmag in the $V$
  band. The spots are assumed to have a temperature lower than that of
  the photosphere by $\Delta T=-500$ K.
\label{fig:spots}}
\end{figure}

We used the method described in \citet{Maxted:2011:w41} to look for
periodic variations due to spots in the lightcurves of WASP-6 from the
WASP archive \citep{Pollacco:2006:wasp}. Data from 3 observing seasons
were analysed independently. The lightcurves typically contain
$\sim$4500 observations with a baseline of 200 nights. From the
projected equatorial rotation velocity of WASP-6 and its radius
\citep{Doyle:2013:wasp_sppars} we estimate that the rotation period is
$16 \pm 3$\, days. There are no significant periodic variations in
this range in any of the WASP lightcurves. To estimate the false alarm
probability of any peaks in the periodogram we use a bootstrap Monte
Carlo method. The results of this anaysis can also be used to estimate
an upper limit of 2 mmag to the amplitude of any periodic variation in
these lightcurves. Therefore, $\epsilon$ is constrained to be less
than the implied peak-to-peak amplitude, $|\epsilon| < 4$ mmag. While
this constraint is valid only at the time the discovery light curve
was taken, lacking any other constrains we will take this value as our
upper limit. In order to estimate $f(\lambda)$ we make use of the high
resolution Phoenix synthetic stellar spectra computed by
\citet{Husser:2013}. We assume $T=5400$ K, $\Delta T = -500$ K, and
other stellar parameters to be the closest available in the models
grid to those presented in \citet{Gillon:2009a}. The resulting
expected maximum value for $\delta k / k$ given the constrains on the
rotational modulation afforded by the WASP-6 discovery light curve is
presented in Figure~\ref{fig:spots}. As can be seen, the change in
$\delta k / k$ induced by starspots over the wavelength range of our
spectrum is expected to be $< 5\times10^{-4}$. This is more than one
order of magnitude less than the change we in $\delta k / k$ we infer
from our observations (see Figure~\ref{fig:transpec_dataonly}), and
thus we conclude that the observed transmission spectrum is not
produced by unocculted spots.

\section{The Transmission Spectrum: Analysis}
\label{sec:discussion}

The main feature of the transmission spectrum shown in
Figure~\ref{fig:transpec_dataonly} is a general sloping trend with
$R_p/R_*$ becoming smaller for longer wavelengths. The general trend
is broken by the two redmost datapoints that could be indicating the
presence of a source of opacity in that region, but the error bars of
the extreme points are large, as the measurements there are naturally
more uncertain because the spectrograph efficiency changes drops
rapidly at the red end of the spectrum and this region of the spectrum
can be badly affected by variations in night sky emission and telluric
absorption. There are no indications of the broad features expected at
the resonance doublets of Na I at 589.4 nm or K I at 767 nm. To make
the statements above quantitative, we compare our measured
transmission spectrum with the clear atmosphere models computed by
\citet{Fortney:2010a}. We scale the models that have a surface gravity
of $g=10$ m/s$^2$ to match the measured surface gravity of WASP-6b
\citep[$g=8.71$ m/s$^2$,][]{Gillon:2009a} by scaling the spectral
features from the base level by $10/g$. We do not have an absolute
reference to be able to place the 10 bar level such as could be
provided by observations at infra-red wavelengths, which in the
\citet{Fortney:2010a} models is set to 1.25 $R_J$, and so we fit for
an overall offset in the $y$-axis. In other words, our measured
transmission spectrum will be able to discriminate on the shape of the
models but will provide no independent information on the absolute
height in the atmosphere where the features are formed. Given that the
equilibrium planet temperature assuming no albedo and full
redistribution between the day and night sides is $T_{\rm eff} =
1194^{+58}_{-57}$ K \citep{Gillon:2009a} we will compare our
measurements with the $T=1000$ and $T=1500$ K models of
\citet{Fortney:2010a}. The $T=1000$ K model has Na and K as the main
absorbers, while the $T=1500$ K model also displays the effects of
partially condensed TiO and VO resulting in a very different
transmission spectra.

In addition to clear atmosphere models we fit our data to a pure
scattering spectrum as given by \citet{lecavelier:2008:rayleigh} for a
scattering cross section $\sigma = \sigma_0(\lambda/\lambda_0)^\alpha$,

\begin{equation}
\frac{d\,R_p}{d\,\ln \lambda} = \alpha H_c = \frac{\alpha k_B T}{\mu g},
\end{equation}

\noindent where $H_c$ is the scale-height of the particles producing
the scattering, which we assume to be equal to the gaseous scale height
$H=k_B T / \mu g$, although condensates producing scattering can
have smaller scale heights than the gas, $H_c \sim H/3$, unless they
are very well mixed vertically \citep{Fortney:2005:condensates}. In
the case of pure scattering we will fit for two parameters to match to
our observed spectrum, the combination $\xi = \alpha T$, and a
zero-point offset. We can then interpret the value of $\xi$ assuming
Rayleigh scattering $\alpha=-4$ or the expected values of the
equlibrium temperature for the atmosphere of WASP-6b.

\begin{figure*}
\plotone{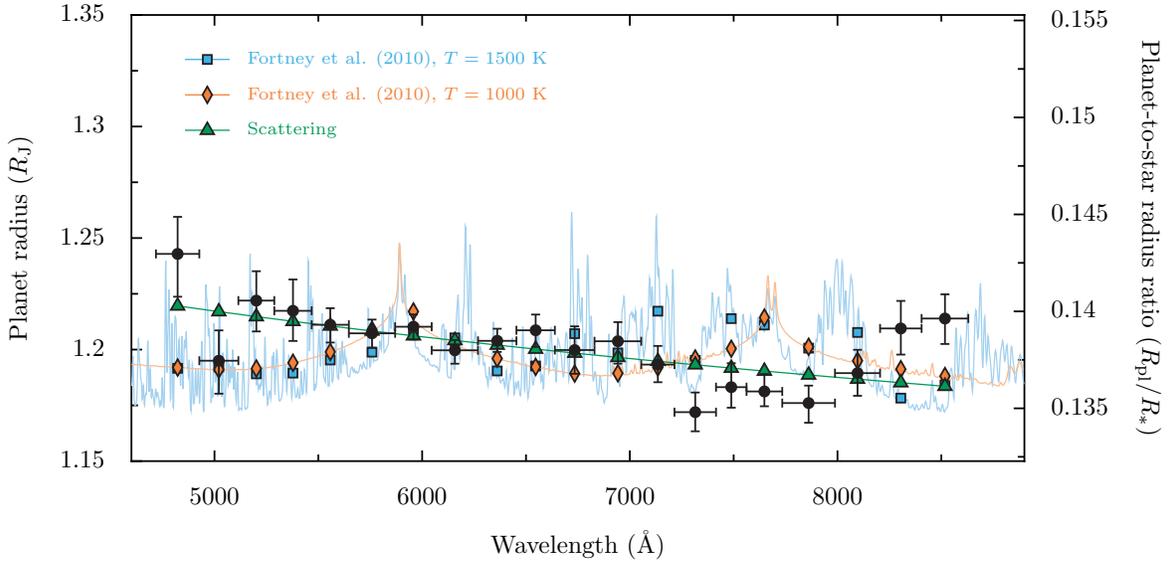}
\caption{Transmission spectrum of WASP-6b along with various models. Black 
dots with error bars indicate our measurements, while blue squares indicate the binned model of 
\cite{Fortney:2010a} with $T_{\rm{eq}} = 1500$ K and red diamonds indicate the binned model using 
$T_{\rm{eq}} = 1500$ K. The green-line and triangles indicate the best-fit line for a scattering 
model, which is the favored model in this case.
\label{fig:transpec}}
\end{figure*}

Along with the transmission spectrum, Figure~\ref{fig:transpec} shows
the results of fitting the models to our observed transmission
spectrum. It is clear to the eye that the best fit is given by the
pure scattering model, with the clear atmosphere models giving
considerably worse fits. The clear atmosphere models fail to give a
better match to the spectrum due to the lack in the latter of evidence
for the broad features expected around Na and K. The AIC for the
scattering model assuming Gaussian noise given by the known error bars
gives $-115.3$, while the values for the $T=1000$ and $T=1500$ clear
atmosphere models are $-97.2$ and $-90.9$ respectively, providing a
very significant preference for the scattering model. A $\chi^2$
analysis gives a $p$-value of 0.04 for the pure scattering model
($\chi^2 = 30$ for 18 degrees of freedom), while the probabilities for
the data being produced by either of the clear atmosphere models are
exceedingly small ($\chi^2 =80$ and $106$ with 19 degrees of freedom
for the $T=1000$ and 1500 models, respectively, giving $p$-values
$<10^{-8}$ for both). Based on the numbers above, only the scattering
model is viable, but these analyses ignore potential correlation of
the errors in the wavelength dimension. Looking at the light curves in
Figure~\ref{fig:transits} one can see that in some cases features in
the light curves repeat between adjacent wavelength channels. In order
to assess the potential impact of correlations in the wavelength
direction, we compute the partial autocorrelation function (PACF) for
the residuals in the wavelength dimension. Denoting the residuals of
the transit fits shown in Figure~\ref{fig:transits} by $r_{il}$, where
$i$ indexes time and $l$ the wavelength channel, we compute the PACF
for the 20 vectors $(r_{1l},r_{2l},\ldots,r_{nl})$ with
$l=1,\ldots,20$ and $n=91$. The PACF has one significant component at
lag 1, which shows that the residuals have indeed correlations with
the adjacent channels that are suggestive of an AR(1) correlation
struture\footnote{AR(p) denotes an ARMA(p,0) process}. This does not
imply that the $(R_p/R_*)_\lambda$ values will necessarily have such
correlation, but we should check how the models fare when including
such potential correlation structure in the fits.

We performed fits then of all three models including an AR(1)
component, to see if it gave a significantly better model as gauged by
the information criteria listed in \S~\ref{ssec:model_sel}. The
scattering model does not need an additional AR(1) component, while
for the clear atmosphere models including correlation gives a
significantly better fit. This should come as no surprise, as the
clear atmosphere models give a poor fit to start with, and
including an extra AR(1) component can effectively model some of the
residual structure. But even after accounting for potential
correlation structure on them, the scattering model is significantly
better than clear atmosphere models. We conclude therefore that our
measured transmission spectrum is most consistent with a featureless
sloped spectrum and does not present significant evidence for the features
predicted by clear atmosphere models even if trying to account for the
differences between the model and the observations with correlated
errors with short lags between the wavelength channels as suggested by
the residuals in the light curves.

The best-fit value for $\xi$ is $\xi = -10670 \pm 3015$. If we fix the
temperature to the equilibrium value given by \citet{Gillon:2009a},
then we would infer $\alpha = -9 \pm 2.5$ and, inversely, when
assuming Rayleigh scattering we'd infer a temperature of
$T=2667\pm750$ K. The inferred values for $\alpha$ and $T$ are
consistent within 2$\sigma$ from the values for Rayleigh scattering
and the equilibrium temperature $T=1194^{+58}_{-57}$ given by \citet{Gillon:2009a}, but the
uncertainties are too large to allow any further conclusions,
especially when considering the additional uncertainty in the
scale height assigned to the material responsible for the scattering.

\section{Discussion and Conclusions}
\label{sec:conclusions}

We have measured the optical transmission spectrum for WASP-6b in the
range $\approx$ 480 to 860 nm via differential spectrophotometry using
7 comparison stars with the IMACS spectrograph on Magellan. By
modeling the systematic effects via a principal component analysis of
the available comparison stars and a white-noise model for the noise,
we are able to achieve light curve with residuals of order $\approx
0.8$ mmag in 200 nm channels per 140 sec exposure, and $\approx 0.5$ mmag
in the summed (white-light) light curve. In order to take into account
possible remaining trends particular to the target star and the
correlated structure of the noise we probe the appropriateness of both
short (ARMA(p,q)) and long ($1/f$ ``flicker'' noise) stochastic
process, making use of well established information criteria to select
the model most appropriate for our particular observations which
turned out to be the $1/f$ model. We believe it is fundamental to
carry out a residual analysis for each particular observation. Lacking
a detailed physical model for a given correlation structure it should
be the data that select which is the most appropriate for a given
observation. With the $1/f$ model the inferred white noise components
are $\approx$ 1--3 times the expected Poisson shot noise
($\sigma_w=0.16$ mmag per 140 sec exposure for the white light curve and $\sim 0.6$ mmag
per 140 sec exposure for the 200nm channels).

The measured spectrum has a general trend of decreasing planetary size
with wavelength, and does not display any evident additional features.
We fit our transmission spectrum with three different models: two
clear transmission spectra from \citet{Fortney:2010a} and a spectrum
caused by pure scattering. Our main conclusion is that the
transmission spectrum of WASP-6b is most consistent with that expected
from a scattering process that is more efficient in the blue. In
addition, the spectrum does not show the expected broad features due
to alkali metals expected in clear atmosphere models which give
significantly less satisfactory description of our data, even when
allowing for the errors to be correlated between different wavelength
bins.

We conclude that the spectrum is most consistent with a featureless
spectrum that can be produced by scattering. The potentially prominent
role of condensates or hazes in determining the transmission spectra
of exoplanets has been apparent from the very first measurement \citep{Charbonneau:2002:hd209}, and
our transmission spectrum of WASP-6b is in line with what seems to be
building trend for transmission spectra with muted features in the
optical. Higher resolution observations around the alkali lines for
WASP-6b will be valuable to see if they remain at detectable levels
over the mechamism that is veiling the very broad lines that are
expected for clear atmospheres. 
We note that the expected equilibrium temperature for WASP-6b is
similar to that of HD~189733b, so it may be the case that a similar
obscurer is acting in both systems.

Our work adds a new instrument (IMACS) to the rapidly increasing set
of ground-based facilities that have been succesfully used to probe
exoplanetary atmospheres.  The constraints that can be obtained using
ground-based facilities is a powerful complement to those possible
from space-based facilities and allow us to access a much broader pool
of systems more representative of the typical brightness of hosts
discovered by ground-based transit surveys. An interesting goal
enabled by this capability will be to probe the transmission spectra
of gas giants with fairly similar surface gravities as a function of
equilibrium temperatures.

\acknowledgements

AJ acknowledges support from FONDECYT project 1130857, BASAL CATA
PFB-06, and the Millenium Science Initiative, Chilean Ministry of
Economy (Nuclei: P10-022-F, P07-021-F).
NE is supported by CONICYT-PCHA/Doctorado Nacional and
MR is supported by FONDECYT postdoctoral fellowship 3120097.
D.K.S. acknowledges support from STFC consolidated grant ST/J0016/1.
J.-M.D. acknowledges funding from NASA through the Sagan Exoplanet
Fellowship program administered by the NASA Exoplanet Science
Institute (NExScI).
A.H.M.J. Triaud is a Swiss National Science Foundation fellow under
grant number PBGEP2-145594.

\bibliographystyle{apj}
\bibliography{refs.bib}

\newpage

\begin{deluxetable}{cllllll}
\tabletypesize{\scriptsize}
\tablecaption{Transit parameters as a function of wavelength\label{tbl-3}}
\tablewidth{0pt}
\tablehead{
\colhead{Wavelength range} & \colhead{$(R_p/R_*)$} & \colhead{$w_1$} &
\colhead{$w_2^\textrm{a}$} & \colhead{$\sigma_w$} &
\colhead{$\sigma_r$} & \colhead{$\sigma_{\rm Poisson}^\textrm{b}$}\\
\colhead{\AA} & \colhead{} & \colhead{} & \colhead{} & 
\colhead{mmag} & \colhead{mmag} & \colhead{mmag} }
\startdata
4718--4927 & $0.1430^{+0.0019}_{-0.0022}$ & $0.9303^{+0.0523}_{-0.1043}$ & 0.2047 & $0.3048^{+0.2399}_{-0.2007}$ & $8.8352^{+0.8051}_{-0.8850}$ & 0.31\\ 
4927--5115 & $0.1375^{+0.0016}_{-0.0017}$ & $0.8705^{+0.0935}_{-0.1616}$ & 0.2016 & $0.5409^{+0.2104}_{-0.2841}$ & $6.5492^{+1.3376}_{-1.6895}$ & 0.29 \\ 
5115--5288 & $0.1406^{+0.0015}_{-0.0016}$ & $0.9127^{+0.0621}_{-0.1148}$ & 0.1955 & $0.7520^{+0.2008}_{-0.2591}$ & $5.8071^{+1.9512}_{-2.0652}$ & 0.30\\ 
5288--5468 & $0.1400^{+0.0016}_{-0.0016}$ & $0.9430^{+0.0427}_{-0.0793}$ & 0.2254 & $0.4151^{+0.1765}_{-0.2394}$ & $6.3147^{+0.9081}_{-1.1713}$ & 0.27\\ 
5468--5647 & $0.1393^{+0.0009}_{-0.0009}$ & $0.8508^{+0.0912}_{-0.1080}$ & 0.2326 & $0.7546^{+0.0771}_{-0.1372}$ & $1.8322^{+2.0657}_{-1.3047}$ & 0.26 \\ 
5647--5870 & $0.1389^{+0.0007}_{-0.0007}$ & $0.8114^{+0.0812}_{-0.0824}$ & 0.2429 & $0.6957^{+0.0492}_{-0.0465}$ & $0.6417^{+0.7686}_{-0.4527}$ & 0.23\\ 
5870--6046 & $0.1392^{+0.0006}_{-0.0007}$ & $0.7259^{+0.0866}_{-0.0897}$ & 0.2467 & $0.6109^{+0.0550}_{-0.0688}$ & $1.3599^{+1.1107}_{-0.9551}$ & 0.25\\ 
6046--6269 & $0.1380^{+0.0005}_{-0.0007}$ & $0.4206^{+0.0897}_{-0.0831}$ & 0.2477 & $0.6289^{+0.0512}_{-0.0456}$ & $0.6554^{+0.8881}_{-0.4690}$ & 0.22\\ 
6269--6454 & $0.1385^{+0.0006}_{-0.0006}$ & $0.5400^{+0.0801}_{-0.0827}$ & 0.2515 & $0.6345^{+0.0452}_{-0.0441}$ & $0.4624^{+0.6751}_{-0.3109}$ & 0.24\\ 
6454--6639 & $0.1390^{+0.0008}_{-0.0009}$ & $0.2265^{+0.1195}_{-0.1152}$ & 0.2718 & $0.6525^{+0.0807}_{-0.1469}$ & $2.2304^{+1.7794}_{-1.6389}$ & 0.24\\ 
6639--6830 & $0.1380^{+0.0012}_{-0.0012}$ & $0.5771^{+0.1581}_{-0.1621}$ & 0.2526 & $0.4817^{+0.1580}_{-0.2070}$ & $4.5171^{+1.2569}_{-1.5051}$ & 0.23\\ 
6830--7055 & $0.1385^{+0.0010}_{-0.0012}$ & $0.3163^{+0.1538}_{-0.1399}$ & 0.2500 & $0.5873^{+0.1152}_{-0.2005}$ & $2.9911^{+1.8578}_{-1.9120}$ & 0.22\\ 
7055--7215 & $0.1373^{+0.0010}_{-0.0009}$ & $0.2642^{+0.1219}_{-0.1172}$ & 0.2484 & $0.6105^{+0.0675}_{-0.0939}$ & $1.8425^{+1.3572}_{-1.1820}$ & 0.27\\ 
7215--7415 & $0.1348^{+0.0010}_{-0.0010}$ & $0.1628^{+0.1236}_{-0.0973}$ & 0.2481 & $0.6886^{+0.0561}_{-0.0648}$ & $1.0756^{+1.2399}_{-0.7853}$ & 0.25\\ 
7415--7562 & $0.1361^{+0.0012}_{-0.0011}$ & $0.3190^{+0.1354}_{-0.1297}$ & 0.2492 & $0.6276^{+0.0957}_{-0.1215}$ & $2.9063^{+1.2681}_{-1.5044}$ & 0.30\\ 
7562--7734 & $0.1359^{+0.0009}_{-0.0008}$ & $0.4066^{+0.1013}_{-0.0971}$ & 0.2493 & $0.6145^{+0.0506}_{-0.0603}$ & $1.0299^{+1.1571}_{-0.7499}$ & 0.31\\ 
7734--7988 & $0.1353^{+0.0009}_{-0.0010}$ & $0.3353^{+0.1135}_{-0.1066}$ & 0.2494 & $0.4310^{+0.0654}_{-0.0670}$ & $2.1595^{+0.7171}_{-0.7261}$ & 0.24\\ 
7988--8205 & $0.1368^{+0.0012}_{-0.0012}$ & $0.3408^{+0.1393}_{-0.1434}$ & 0.2478 & $0.3215^{+0.0903}_{-0.1173}$ & $3.6847^{+0.6986}_{-0.6914}$ & 0.28\\ 
8205--8405 & $0.1391^{+0.0014}_{-0.0013}$ & $0.2065^{+0.1626}_{-0.1316}$ & 0.2471 & $0.3888^{+0.1214}_{-0.1655}$ & $4.7549^{+0.8495}_{-0.9059}$ & 0.31\\ 
8405--8630 & $0.1396^{+0.0012}_{-0.0013}$ & $0.2599^{+0.1584}_{-0.1405}$ & 0.2454 & $0.4683^{+0.0991}_{-0.1517}$ & $4.4703^{+0.9456}_{-0.8350}$ & 0.30\\

\enddata
\tablenotetext{a}{$w_2$ is fixed to the values calculated as
  described in \citet{Sing:2010a} for the stellar parameters
  appropriate for WASP-6. Parameters not shown in this Table are fixed
to the posterior values obtained form the white light curve analysis
shown in Table~\ref{tab:wl_transit}.}
\tablenotetext{b}{Expected Poisson noise level.}
\label{tab:transits}
\end{deluxetable}

\end{document}